\documentclass[lettersize,journal]{IEEEtran}
\usepackage{amsmath,amsfonts}
\usepackage[linesnumbered,ruled,vlined]{algorithm2e}
\usepackage{algorithmic}
\usepackage{array}
\usepackage[caption=false,font=normalsize,labelfont=sf,textfont=sf]{subfig}
\usepackage{textcomp}
\usepackage{stfloats}
\usepackage{url}
\usepackage{verbatim}
\usepackage{graphicx}
\usepackage{newtxtext,newtxmath} 
\def\BibTeX{{\rm B\kern-.05em{\sc i\kern-.025em b}\kern-.08em
    T\kern-.1667em\lower.7ex\hbox{E}\kern-.125emX}}
\usepackage{balance}
\usepackage{cite}
\begin{document}

\title{Joint Cache Placement and Routing in Satellite-Terrestrial Edge Computing Network: A GNN-Enabled DRL Approach}

\author{Yuhao Zheng,
\IEEEmembership{Graduate Student Member,~IEEE,}
Ting You,
\IEEEmembership{Student Member,~IEEE,}
Kejia Peng,
and Chang Liu
\thanks{
Yuhao Zheng is with the School of  Information and Communication Engineering, Beijing University of Posts and Telecommunications, Beijing, 100876, China (e-mail: yuhao\_zheng@bupt.edu.cn).

Ting You and Chang Liu are with Beijing Sport University, Beijing, 100084, China (e-mail: t.you@bsu.edu.cn; c.liu@bsu.edu.cn).

Kejia Peng is with Nanjing University of Science and Technology, Nanjing, 210094, China, and also with Mendeleev University of Chemical Technology of Russia, Moscow, 125047, Russia (e-mail: 230997@muctr.ru).

*Corresponding authors: Yuhao Zheng, Chang Liu
}
}

\markboth{Journal of \LaTeX\ Class Files,~Vol.~14, No.~8, August~2021}%
{How to Use the IEEEtran \LaTeX \ Templates}


\maketitle

\begin{abstract}
In this letter, we investigate the problem of joint content caching and routing in satellite-terrestrial edge computing networks (STECNs) to improve caching service for geographically distributed users. To handle the challenges arising from dynamic low Earth orbit (LEO) satellite topologies and heterogeneous content demands, we propose a learning-based framework that integrates graph neural networks (GNNs) with deep reinforcement learning (DRL). The satellite network is represented as a dynamic graph, where GNNs are embedded within the DRL agent to capture spatial and topological dependencies and support routing-aware decision-making. The caching strategy is optimized by formulating the problem as a Markov decision process (MDP) and applying soft actor-critic (SAC) algorithm. Simulation results demonstrate that our approach significantly improves the delivery success rate and reduces communication traffic cost.
\end{abstract}

\begin{IEEEkeywords}
Satellite-terrestrial network, graph neural network, deep reinforcement learning, content placement.
\end{IEEEkeywords}

\section{Introduction}
\IEEEPARstart{W}{ith} the continuous evolution of 6G technology and satellite communications, the global demand for seamless connectivity and ubiquitous computing services is rapidly increasing \cite{ref1}. To meet the requirements of wide coverage and highly reliable communication, the satellite-terrestrial network (STN) has emerged to provide broad access \cite{ref2}. Meanwhile, mobile edge computing (MEC) has become a key technology for addressing the low-latency needs by deploying computing resources closer to users \cite{ref3}. Based on the integration of STN and MEC, the satellite-terrestrial edge computing network (STECN) has been proposed as a critical architecture for supporting wide-area intelligent services \cite{ref4}. STECN enhances the capacity of terrestrial networks in densely populated urban areas and extends computing services to remote regions, thereby enabling the deployment of latency-sensitive and computation-intensive applications across a broader geographic area \cite{ref5}.

STECN introduces edge computing into satellite-terrestrial architectures to enhance computational responsiveness, transmission efficiency, and real-time task processing in recent years. Esmat \textit{et al.} in \cite{ref6} proposed a slicing scheme using Markov models and restless multi-armed bandits to optimize slice allocation in STECN. Huang \textit{et al.} in \cite{ref7} developed a dual-timescale optimization framework to solve task scheduling and resource slicing problem in STECN. Current research largely focuses on slicing and task scheduling, while edge caching remain underexplored. With rising demand for artificial intelligence (AI) services, satellite caches are shifting from traditional data to AI-related models or images. Pre-cached computational images or function modules in edge caches can be directly utilized to facilitate fast task execution. Central challenge is to dynamically decide what content to cache and when to replace it, considering request and resource states.

Despite the great potential of STECN in enabling global edge intelligence caching services, its highly dynamic network structure and limited onboard resources pose significant challenges. Unlike static terrestrial networks, the fast movement of LEO satellites causes frequent changes in network topology, making it critical to dynamically cache the appropriate AI model versions for different regions at different times. Given the limited onboard storage, it is infeasible to cache all models simultaneously, necessitating trade-offs among task hotspots, content validity, and network conditions. Therefore, there is an urgent need for intelligent mechanisms that integrate network-awareness, temporal modeling, and adaptive decision-making to enable dynamic optimization of caching.

To address the challenges in STECN, we propose a novel learning-based framework that jointly optimizes caching and routing. The network is modeled as a dynamic spatiotemporal graph, where graph neural networks (GNNs) are embedded into the reinforcement learning agent to enable routing representation. The caching decision is formulated as a Markov decision process (MDP) and optimized using deep reinforcement learning (DRL). GNNs capture the structural information for routing, while DRL determines cache placement strategies. This integration enables the system to adaptively learn strategies to improve delivery success rate and reduce traffic.

\section{System Description}\label{SectionII}
In this section, we first describe our STECN system model.

\subsection{Network Model}
We consider STECN where LEO satellites provide caching to support terrestrial user requests. Due to the highly dynamic and periodic nature of LEO satellites, the system operates in discrete time slots indexed by $t \in \{1, 2, \ldots, T\}$. Let $\mathcal{N} = \{1, 2, \ldots, N\}$ denote the set of LEO satellites, where each LEO satellite $n \in \mathcal{N}$ is equipped with a local cache. The LEO satellites form a dynamically changing topology network that evolves periodically with orbital motion. The relative positions of LEO satellites within the same orbital plane remain stable, and stable inter-orbital distances enable connections with nearby LEO satellites in adjacent planes \cite{ref8}. With four inter-satellite links (ISLs), LEO satellites form a grid topology orbital network.

\subsection{Communication Model}
The communication model includes the ISL model and the downlink model.

\subsubsection{Inter-Satellite Link Model}
LEO satellites use laser communication to establish ISLs. The data transmission rate from LEO satellite $n$ to satellite $m$ at time slot $t$ is given by \cite{ref9}:
\begin{equation}
R_{n,m}(t) = \frac{P_n G_n^{\mathrm{TX}} G_m^{\mathrm{RX}} L_{n,m}(t)}{h_1 T^{S} \cdot (E_b/N_0)_{\mathrm{req}} \cdot M},
\end{equation}
where $L_{n,m}(t) = \left(\frac{c}{4\pi H_{n,m}(t) f}\right)^2$ denotes the free space path loss. $c$ is the speed of light, $f$ is the carrier frequency of the ISL, and $H_{n,m}(t)$ is the distance between LEO satellite $n$ and $m$ at time $t$. $P_n$ is the transmission power of LEO satellite $n$, $G_n^{\mathrm{TX}}$ and $G_m^{\mathrm{RX}}$ are the antenna gains of LEO satellites $n$ and $m$, respectively. $h_1$ is the Boltzmann constant, $T^{S}$ is the system noise temperature, $(E_b/N_0)_{\mathrm{req}}$ is the required energy per bit to noise power spectral density, and $M$ is the link margin.

\subsubsection{Downlink Model}
The channel from LEO satellites to ground users is mainly influenced by rain attenuation modeled by Weibull distribution. The power attenuation from LEO satellite $n$ to the ground at time $t$ is given by \cite{ref10}:
\begin{equation}
\|h_{n,0}(t)\|^2 = \frac{G_n^{\mathrm{TX}} G_0^{\mathrm{RX}} \lambda^2}{(4\pi H_{n,0}(t))^2} \cdot 10^{-F^{\mathrm{rain}}/10},
\end{equation}
where $\lambda$ is the carrier wavelength, $G_0^{\mathrm{RX}}$ is the antenna gain of ground user device. $H_{n,0}(t)$ is the distance between LEO satellite $n$ and ground user at time $t$. $F^{\mathrm{rain}}$ is the rain attenuation following Weibull distribution. The maximum transmission rate from LEO satellite $n$ to the ground at time $t$ is given by:
\begin{equation}
R_{n,0}(t) = W \log_2\left(1 + \frac{P_n \|h_{n,0}(t)\|^2}{W \|\sigma\|^2}\right),
\end{equation}
where $W$ is the communication bandwidth of the LEO satellite, $\sigma^S$ is the noise power. We use orthogonal frequency division multiplexing (OFDM) to assign unique subchannels to requests, and to ensure orthogonality to prevent interference.

\subsection{Cache Model}
\subsubsection{Request Model}
The set of contents is defined as $\mathcal{F} = \{1, \ldots, F\}$, where the size of file $f$ is denoted by $\zeta_f$. The popularity distribution of content is characterized by the Zipf distribution. The popularity of the $f$-th ranked content is \cite{ref11}:
\begin{equation}
P_f = \frac{f^{-\alpha}}{\sum_{f=1}^{F} f^{-\alpha}}, \quad \forall f \in \mathcal{F}, \label{eq:zipf}
\end{equation}
where $\alpha$ is Zipf exponent. Let $\mathcal R_n(t) = \{r_{n,1}(t), \ldots, r_{n,F}(t)\}$ be the request set received by LEO satellite $n$ at time $t$, where $r_{n,f}(t)$ represents the number of requests for $f$.

\subsubsection{Content Caching Model}
To describe the caching status of each LEO satellite in the system, we define a content caching matrix. We denote the caching status of the system at time $t$ as $\Psi(t) = \{\Psi_1(t), \Psi_2(t), \ldots, \Psi_N(t)\}$, where $\Psi_n(t) = \{\Psi_{n,1}(t), \Psi_{n,2}(t), \ldots, \Psi_{n,F}(t)\}$ represents the caching vector of the LEO satellite $n$. $\Psi_{n,f}(t) = 1$ indicates that LEO satellite $n$ caches content $f$ in time slot $t$, otherwise, $\Psi_{n,f}(t) = 0$.

\subsection{Success Model}
\subsubsection{Delivery Delay}
The total delay for delivering content $f$ from LEO satellite $n$ to the ground user can be expressed as:
\begin{equation}
D_{n,f}(t) = \frac{\zeta_f}{R_{n,0}^{ST}(t)}+(1-\Psi_{n,f})\cdot\sum_{p,q\in e(n,m)}{\frac{\zeta_f}{R_{p,q}^{S}(t)}}
\end{equation}
where $e(n,m)$ is the transmission path in the ISL. If the LEO satellite $n$ caches $f$, there is no need for ISL transmission. Otherwise, the request need to find LEO satellite $m$ that meets the satisfaction $\Psi_{m,f}(t)=1$.

\subsubsection{Success Rate}
Reliability is a fundamental indicator of system performance. A key threat to reliability is request expiration, which occurs when a request fails to complete within its tolerable latency threshold $D_f^{\text{thres}}(t)$. The number of discarded request at time slot $t$ is calculated as:
\begin{equation}
K(t) = \sum_{n \in \mathcal{N}} \sum_{r_{}(t) \in \mathcal{R}_n(t)} \mathbb{D}[D_{n,f}(t) >D_f^{\text{thres}}(t)]
\end{equation}
where $\mathbb{D}[\cdot]$ is an indicator function that equals 1 if the condition is true, and 0 otherwise. The success rate is the ratio of successful requests to the total requests across all LEO satellites:
\begin{equation}
S(t) = 1 - \frac{K(t)}{\sum_{n \in \mathcal{N}} |\mathcal{R}_n(t)|}.
\label{success}
\end{equation}

\subsection{Traffic Model}
We construct traffic model from two perspectives as follows.

\subsubsection{Request Response Traffic}
If  the request content is not locally cached, the content must be fetched via ISL. The total transmission traffic at LEO satellite $n$ in time slot $t$ is given by:
\begin{equation}
{Tr}_n^{\text{req}}(t) = \sum_{r(t) \in Q_n(t)} \left( \zeta_f + \left(1-\Psi_{n,f}(t)\right)\cdot\sum_{e\in e(n,m)}   \zeta_f \right)
\end{equation}

\subsubsection{Cache Update Traffic}
The cache states dynamically change. Let $\Psi_{n,f}(t-1)$ denote the cache state in the previous time slot. If $\Psi_{n,f}(t-1)\neq\Psi_{n,f}(t)=1$, it indicates that content $f$ is newly cached. The cache update traffic at LEO satellite $n$ in time slot $t$ can be expressed as:
\begin{equation}
{Tr}_n^{\text{update}}(t) = \sum_{f=1}^F \mathbb{D}[\Psi_{n,f}(t-1)\neq\Psi_{n,f}(t)=1] \cdot \zeta_f
\end{equation}

The total traffic of the system at time slot $t$ is defined as:
\begin{equation}
{Tr}(t) = \sum_{n=1}^N \left( {Tr}_n^{\text{req}}(t) + {Tr}_n^{\text{update}}(t) \right).
\label{traffic}
\end{equation}

\subsection{Objective}
In our work, the objective is to jointly maximize the success rate and minimize the communication traffic overhead. Therefore, the optimization problem is defined as:
\begin{equation}
\text{(P1)}: \max_{\pi} \lim_{T \to \infty} \frac{1}{T} \sum_{t=1}^{T} E \left[ c(t) \mid \pi \right]
\end{equation}
where the reward function is formulated as:
\begin{equation}
c(t) = \lambda_1 \cdot S(t) - \lambda_2\cdot\frac{{Tr}(t)}{{Tr}_{max}},
\end{equation}
where $\lambda_1$, $\lambda_2$ are weight coefficients. ${Tr}_{max}$ is the maximum update traffic as a normalization factor.

\section{Graph-Temporal SAC-Based Caching and Routing Optimization}\label{SectionIII}
This section introduces our proposed routing and caching optimization method for dynamic LEO satellite networks.

\subsection{Overview of the GT-SAC Framework}
To address the challenges posed by rapidly changing LEO satellite topologies and user requests, we design a graph-temporal SAC (GT-SAC) algorithm. This method models the LEO satellite network as a dynamic graph structure and employs GNN \cite{ref12} to extract spatial dependencies. Temporal fluctuations in user demands and network states are captured via a sequential modeling mechanism. These spatiotemporal features are then fed into a SAC-based agent to jointly optimize routing paths and caching decisions. By coupling graph-based representation learning with policy optimization, GT-SAC effectively captures the structural and temporal dynamics of the LEO satellite network, enabling adaptive and efficient decision-making.

\subsection{Spatiotemporal Representation of Satellite Network}
\subsubsection{Graph Construction}
We model the LEO satellite network at time slot $t$ as a dynamic undirected graph $\mathcal{G}_t = (\mathcal{V}_t, \mathcal{E}_t)$, where $\mathcal{V}_t=\mathcal{N}_t$ represents LEO satellites and each edge $\mathcal{E}_t$ denotes active ISLs. Each LEO satellite is represented by a feature vector $\mathbf{x}_n^t$, which incorporates its 3D spatial coordinates $(\phi_n^t, \lambda_n^t, h_n^t)$, the corresponding regional one-hot encoding $l_n^t$, its content request matrix $\mathcal{R}_n(t)$, and cache state $\Psi_n(t) \in \{0,1\}^{|\mathcal{F}|}$. These features are aggregated into form a matrix $\mathbf{X}_t \in \mathbb{R}^{|\mathcal{V}_t| \times d}$. Edge features, including distance $d_{ij}^t$ and bandwidth $b_{ij}^t$ are stored in an edge feature matrix $\mathbf{E}_t \in \mathbb{R}^{|\mathcal{\mathcal{E}}_t| \times d}$, representing real-time communication quality.

\subsubsection{Message-Passing-Based Routing Representation}
To enable distributed routing decision-making in the dynamic LEO satellite graph, we adopt a spatiotemporal message-passing neural network that explicitly incorporates edge features. Each LEO satellite node $i$ is initialized with a feature vector $\mathbf{x}_i^t$, each edge $e_{ij}$ is associated with a feature vector $\mathbf{e}_{ij}^t$. The node embeddings are updated through iterative message passing. The node embeddings are updated through iterative message passing. At each iteration $t$, the hidden representation $\mathbf{h}_i^{(t)}$ is updated as:
\begin{equation}
\mathbf{h}_i^{(t+1)} = \sigma\left( \sum_{j \in \mathcal{N}(i)} f_\text{msg}(\mathbf{h}_i^{(t)}, \mathbf{h}_j^{(t)}, \mathbf{e}_{ij}^t) \right),
\label{eq:gnn-routing}
\end{equation}
where $f_\text{msg}(\cdot)$ is a learnable message function that fuses neighbor node features and link characteristics, and $\sigma(\cdot)$ is a non-linear activation function. After $T$ iterations, final embedding $\mathbf{h}_i^{(T)}$ encodes both multi-hop topological structure and communication quality used for routing and DRL caching decisions.

\begin{algorithm}[!t]
\caption{GT-SAC Cache Optimization Algorithm}
\label{alg:cache_opt_SAC}
\KwIn{User request set $\mathcal{R}(t)$, previous cache state $\Psi(t{-}1)$, communication state $R_n(t)|_{n \in \mathcal{N}}$}

\KwOut{Optimized caching strategy $\pi$}

Initialize network $\pi_{\phi}$, $Q_{\theta_1}$, $Q_{\theta_2}$, $V_{\psi}$, $V_{\psi'}$, $M_R$, $M_b$.

\For{episode = 1 to $Epsiodes$}{
    Initialize environment and initial state $s_0$

    \For{t = 1 to $T$}{
        \tcp{Graph Construction}
        Construct $\mathcal{G}_t = (\mathcal{V}_t, \mathcal{E}_t)$ based on topology and requests.

        Extract node feature.

        Encode global state $s_t$ by GNN.

        \tcp{Experience Storage}
        Select action $a_t$ based on policy $\pi_{\phi}$.

        Execute $a_t$, update cache state $\Psi(t)$.

        \tcp{Routing via Message Passing}
        Message passing to get embedding by Eq.~(\ref{eq:gnn-routing}).
        
        Route requests over $\mathcal{G}_t$ based on embeddings.

        Obtain $S(t)$ and $Tr(t)$ by Eq.~(\ref{success},\ref{traffic}).

        Observe environment reward $r_t$ and state $s_{t+1}$.

        Store transition $(s_t, a_t, r_t, s_{t+1})$ in buffer $M_R$.

    \tcp{SAC Optimization Phase}

    Sample $M_b$ from buffer $M_R$.

    Calculate Q-value by Eq.(\ref{eq:Q_target});
    
    Update Q-Critic network by Eq.(\ref{eq:Q_loss});
    
    Calculate V-value by Eq.(\ref{eq:V_target});
    
    Update V-Critic network by Eq.(\ref{eq:V_loss});
    
    Update actor network by Eq.(\ref{eq:actor_loss});
    
    Soft update target V-Critic network by Eq.(\ref{eq:soft_update});
    }
}
\end{algorithm}

\begin{figure}[!t]
\centering
\includegraphics[width=3.45in]{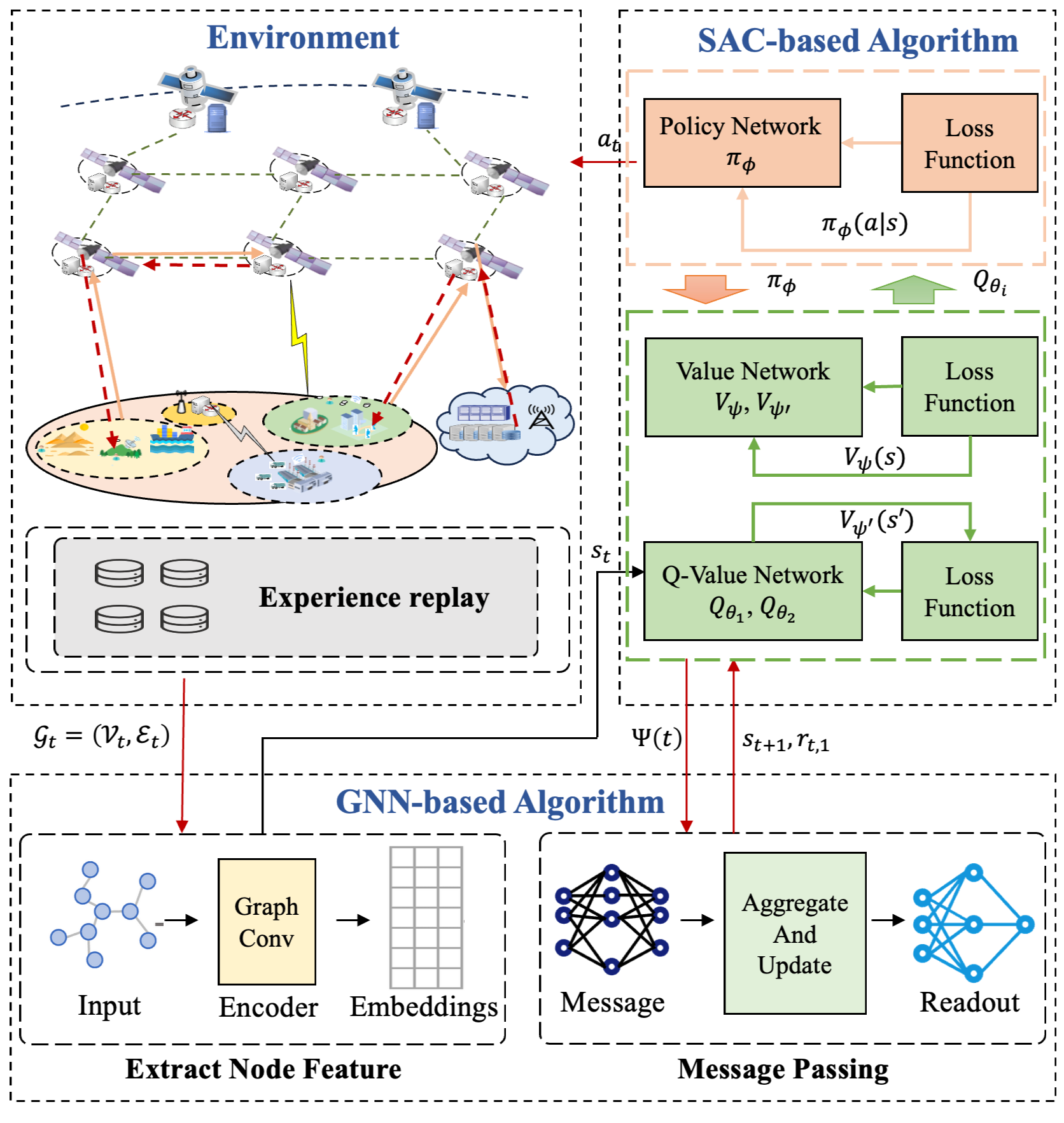}
\caption{GT-SAC algorithm framework.}
\label{Algorithm_Architecture}
\end{figure}

\subsection{SAC-Based Joint Routing and Caching Optimization}

\subsubsection{MDP Formulation}
Our objective is to achieve minimization of the traffic cost and maximization of the success rate. Since the solution is influenced by current state but independent of previous states, the problem can be formulated as MDP. The state space, action space, and cost function are defined as follows.

$\bullet$ \textit{State:} Let $S$ denote the set of possible states. The composite state at timeslot $t$, represented as $s_t \in S$, is defined as $s_t = \{R(t),\Psi(t-1),\mathcal{R}_n(t)|_{n\in\mathcal{N}}\}$.

$\bullet$ \textit{Action:} execute computing task action. The action is denoted as $a_t \in A$ and can be described as $a_t = \{\Psi(t)\}$.

$\bullet$ \textit{Reward:} The objective of optimization is to comprehensively the state of LEO satellites to make optimal caching placement decisions. Based on the optimization goal of maximizing reward, we define the reward at time slot $t$ as $r_t =c(t)$.

\subsubsection{Algorithm Design}
We propose a solution based on DRL, utilizing the soft actor-critic (SAC) \cite{ref13} to train the policy. We considered several common algorithms, such as deep Q-Network (DQN). However, DQN performs poorly in high-dimensional action spaces, leading to increased computational complexity and slower convergence. SAC is based on maximum entropy, which maximizes both cumulative rewards and policy entropy to encourage exploration. The steps are as follows.

$\bullet$ \textit{Initialization:} We initialize network $\pi_{\phi}$, $Q_{\theta_1}$, $Q_{\theta_2}$, $V_{\psi}$ and $V_{\psi'}$. Initialize experience replay buffer $M_r$, mini-batch size $M_b$.

$\bullet$ \textit{Experience storage:} The agent uses the actor network to compute the action probabilities $\pi_{\phi}(a | s_t)$. $a_t$ is sampled based on these probabilities and obtain state $s_{t+1}$ and reward $r_{t}$. $(s_t, a_t, r_t, s_{t+1})$ is stored in $M_r$.

$\bullet$ \textit{Parameter Update:} The actor and critic networks update their parameters by randomly sampling $M_b$ from $M_r$.

\textit{a. Q-critic network update}: The smaller Q-value between $Q_{\theta_1}$ and $Q_{\theta_2}$ is considered the true $Q$-value. $\theta_i$, where $i \in \{1,2\}$, are updated by minimizing the loss function:
\begin{equation}
    L_Q(\theta_i) = \mathbb{E}_{M_b} \left[ (Q_{\theta_i}(s_t, a_t) - y_t^Q)^2 \right], \quad i=1,2
    \label{eq:Q_loss}
\end{equation}
where $y_t^Q$ is the Q value, computed as:
\begin{equation}
    y_t^Q = r_t + \gamma V_{\psi'}(s_{t+1}).
    \label{eq:Q_target}
\end{equation}
where $\gamma$ is the discount rate, $V_{\psi'}(s_{t+1})$ is the output of the target V-critic network.

\textit{b. V-critic network update}: 
The state value is estimated using entropy-regularized formulation, value of V-critic network is:
\begin{equation}
    y_t^V = \mathbb{E}_{a_t \sim \pi} \left[ \min_{i=1,2} Q_{\theta_i}(s_t, a_t) - \alpha \log \pi(a_t | s_t) \right].
    \label{eq:V_target}
\end{equation}

V-critic network is updated by minimizing the loss function:
\begin{equation}
    L_V(\psi) = \mathbb{E}_{M_b} \left[ (V_{\psi}(s_t) - y_t^V)^2 \right].
    \label{eq:V_loss}
\end{equation}

\textit{c. Actor update}: The policy network aims to maximize both rewards and entropy. The policy loss function, derived from the Kullback-Leibler (KL) divergence, simplifies to:
\begin{equation}
    J_{\pi}(\phi) = \mathbb{E}_{M_b, a_t \sim \pi} \left[ \alpha \log \pi(a_t | s_t) - \min_{i=1,2} Q_{\theta_i}(s_t, a_t) \right].
    \label{eq:actor_loss}
\end{equation}
where $\phi$ are updated by determining the gradient of $J_{\pi}(\phi)$.

\textit{d. Soft update}: SAC employs soft updates to gradually update target network. The target V-critic network is updated as follows:
\begin{equation}
    \psi' \leftarrow \tau \psi + (1 - \tau) \psi',
    \label{eq:soft_update}
\end{equation}
where $\tau$ is hyperparameter controlling the update rate. The DRL model is pre-trained on ground servers with high-performance hardware. After convergence, the trained model is deployed to LEO satellites. The steps of the proposed algorithm are summarized in Algorithm \ref{alg:cache_opt_SAC} and Fig.\ref{Algorithm_Architecture}. 

\section{Simulation Results}\label{SectionIV}
In this section, we perform extensive simulations to evaluate the performance of our proposed algorithm.

\subsection{Simulation Setup}
During the simulation process, we used the PyTorch framework to construct our software environment. We established a Walker constellation consisting of 144 LEO satellites, evenly distributed across 12 circular orbits, with 12 LEO satellites per plane. The LEO satellite altitude is set to 1000km, and the orbital plane inclination is 60°. We utilized the satellite tool kit (STK) to obtain latitude and longitude for LEO satellites. The LEO satellite transmission power is assumed to be 5W, and the total system noise temperature is 25dBK. For ISLs, the transmission and reception antenna gains of LEO satellites are both set to 20dB. The required energy per bit to noise density ratio is 9.6dB, and the link margin is 2500km. We set up 6 cache contents, each with a size of 100MB. The timeslot is set to 5 seconds. Numerical results are obtained after 500 training episodes. We adopt 2 propagation layers with hidden dimension 32 to implement the message function followed by ReLU activation. We compare our proposed scheme with baseline methods: SAC, popular-content-first (PCF), and cloud scheme. The SAC scheme employs the same reinforcement learning framework but without GNN-based routing, retrieving content only from neighboring LEO satellites. The PCF scheme caches the most frequently requested contents based on request matrix. The cloud scheme retrieves content from the ground cloud center.

\subsection{Simulation Results}

\begin{figure*}
	\centering
	\subfloat[\textnormal{Different algorithms}] 
	{
			\begin{minipage}[b]{.31\textwidth}
					\centering
					\includegraphics[width=\textwidth]{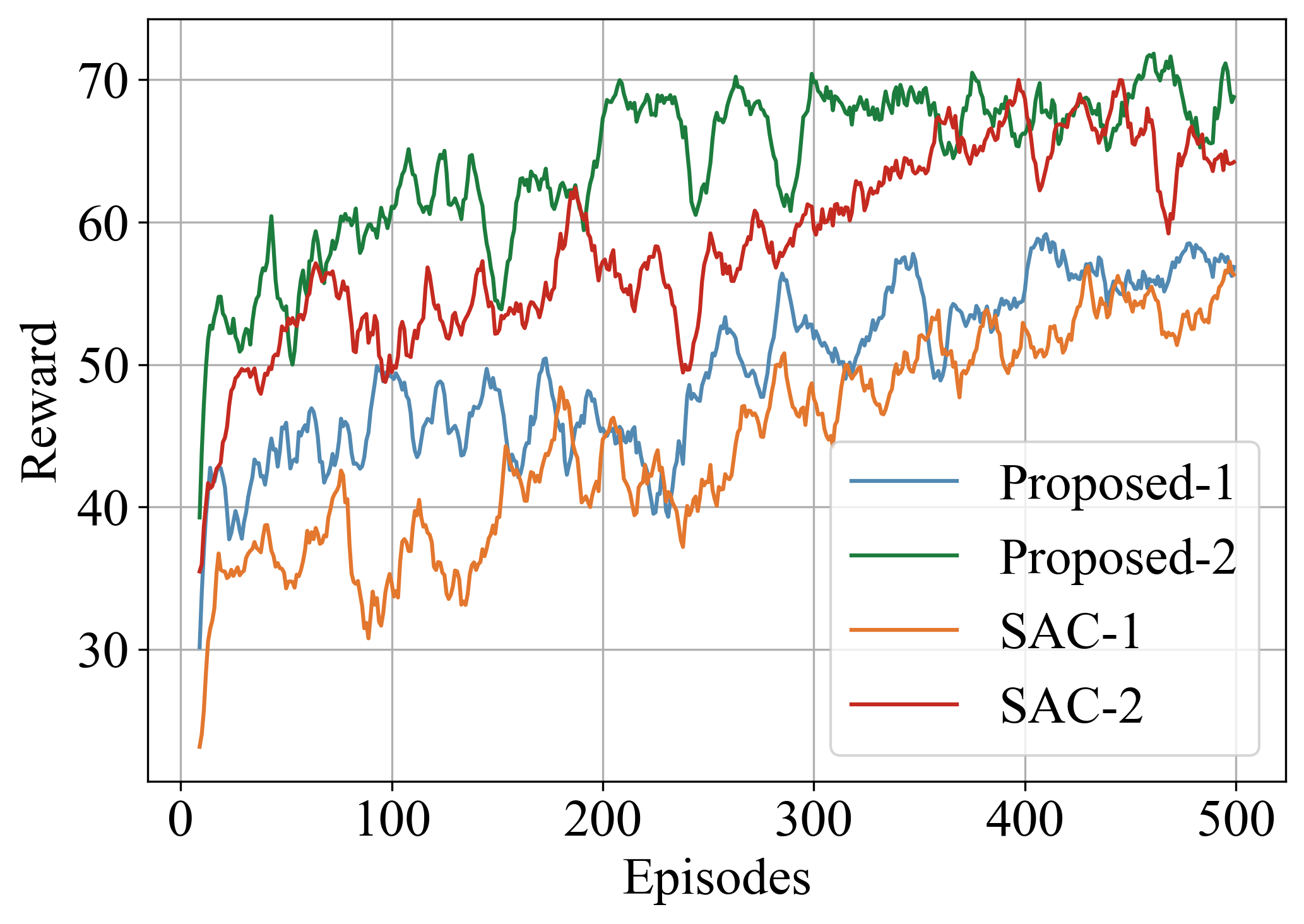}
					\label{Algorithm_Compare}
                    \vspace{-5mm} 
			\end{minipage}
		}
	\subfloat[\textnormal{Different learning rates}] 
	{
			\begin{minipage}[b]{.31\textwidth}
					\centering
					\includegraphics[width=\textwidth]{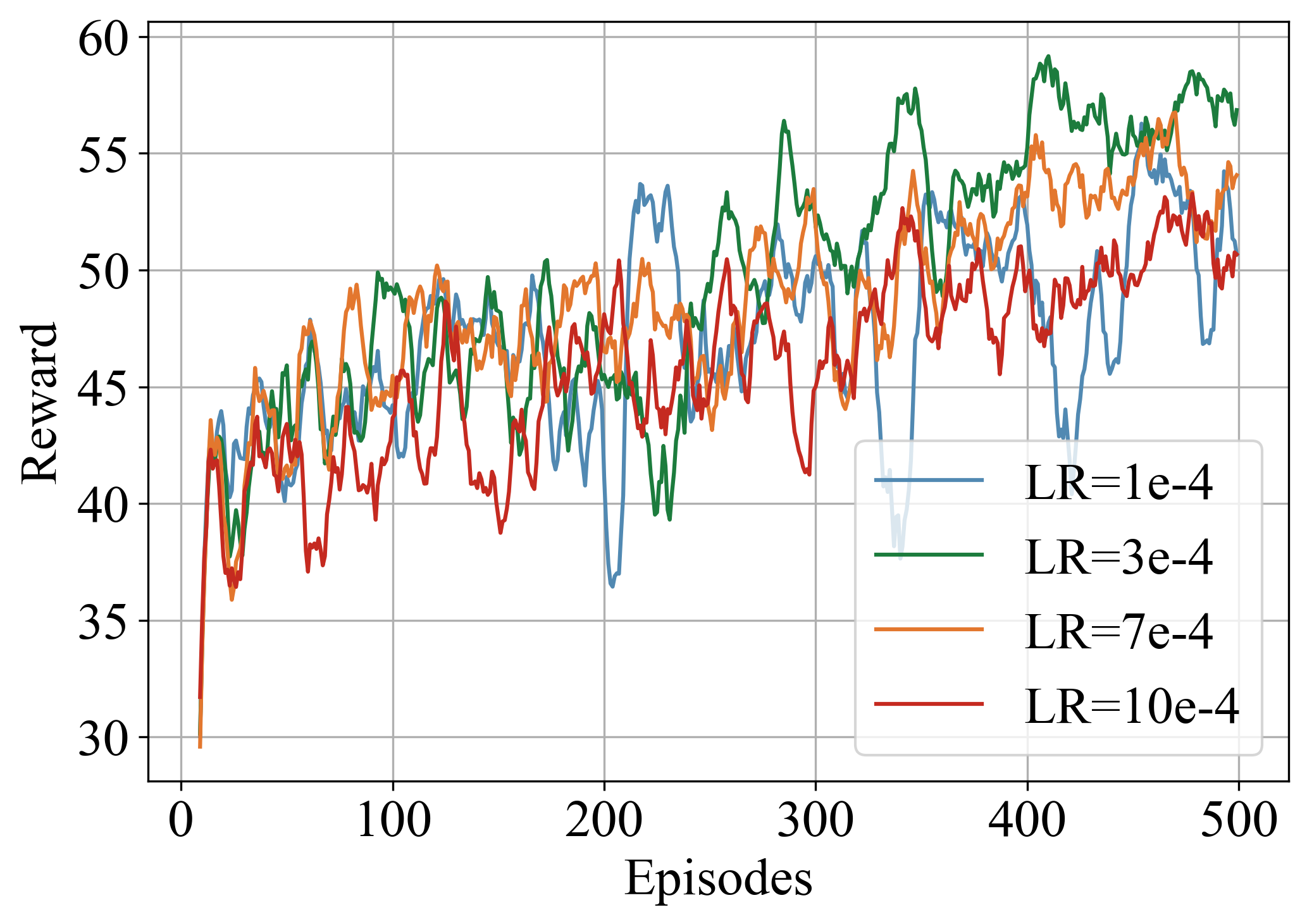}
					\label{LearningRate_Compare}
                    \vspace{-5mm} 
			\end{minipage}
		}
    \subfloat[\textnormal{Different batch sizes}] 
	{
			\begin{minipage}[b]{.31\textwidth}
					\centering
					\includegraphics[width=\textwidth]{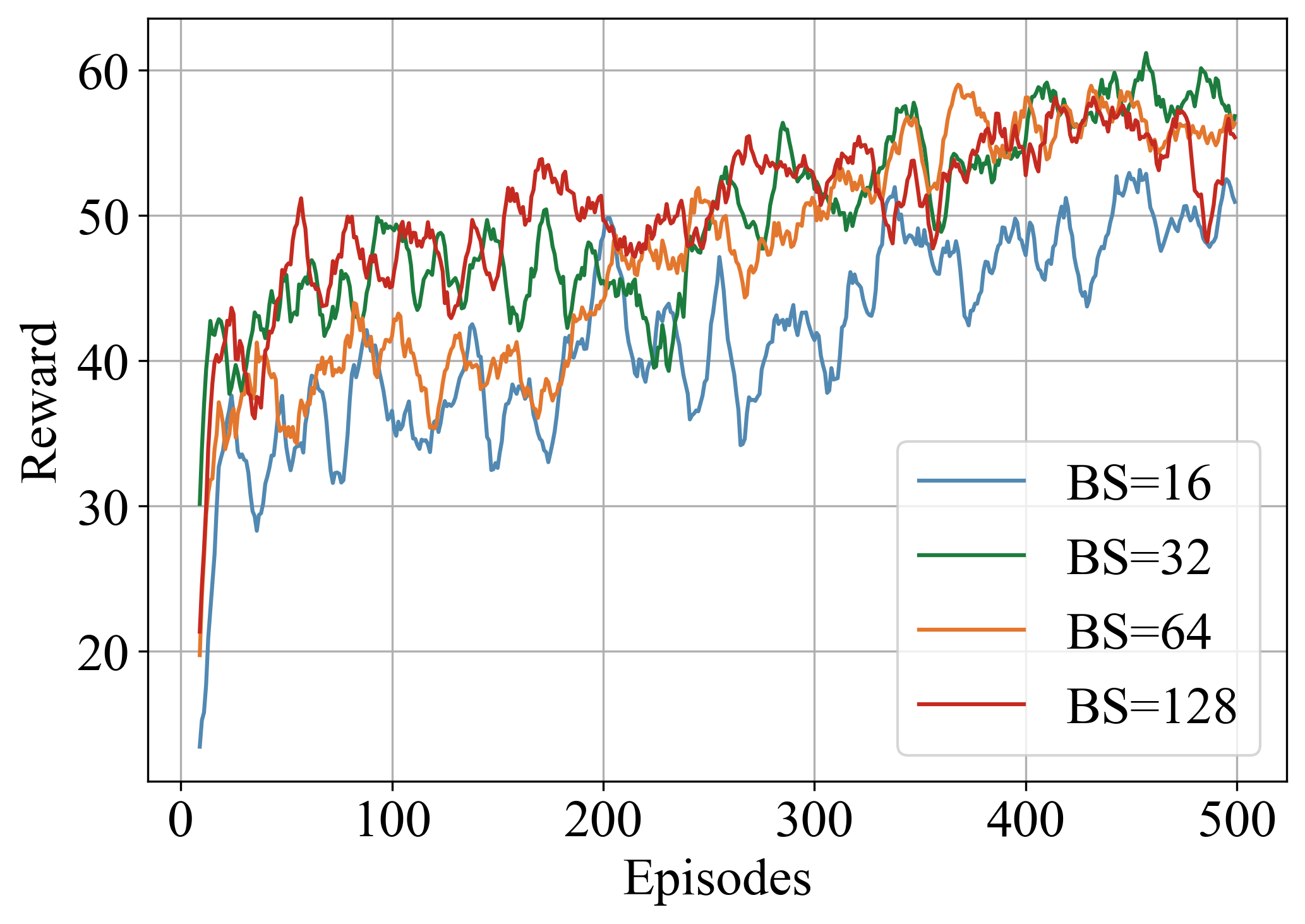}
					\label{BatchSize_Compare}
                    \vspace{-5mm} 
			\end{minipage}
	}

	\caption{Convergence performance of DRL-based algorithms.}
    \label{DRL_Performance}
\end{figure*}

We first conducted simulations to validate the effectiveness of the proposed method. In Fig.\ref{DRL_Performance}, we comprehensively evaluate the DRL algorithms and the hyperparameter settings of the proposed algorithm. In Fig.\ref{Algorithm_Compare}, under the same training conditions, the proposed algorithm demonstrates the best performance as training progresses under different caching numbers, eventually achieving stable convergence. In Fig.\ref{LearningRate_Compare}, we compare the training performance of the proposed algorithm under different learning rates. When the learning rate is set to 0.0003, the reward value increases rapidly in the early training stages, and the algorithm converges faster. Therefore, we set LR at 0.0003 for subsequent experiments. In Fig.\ref{BatchSize_Compare}, we present the training performance of the proposed algorithm under different batch sizes. When we set batch size to 32, the algorithm exhibits better convergence and stability. Overall, the proposed algorithm outperforms the other three algorithms, and it achieves the best training performance when learning rate is 0.0003 and batch size is 32.

\begin{figure}
	\centering
	\subfloat[\textnormal{Cache number is 1}] 
	{
			\begin{minipage}[b]{.25\textwidth}
					\centering
					\includegraphics[width=\textwidth]{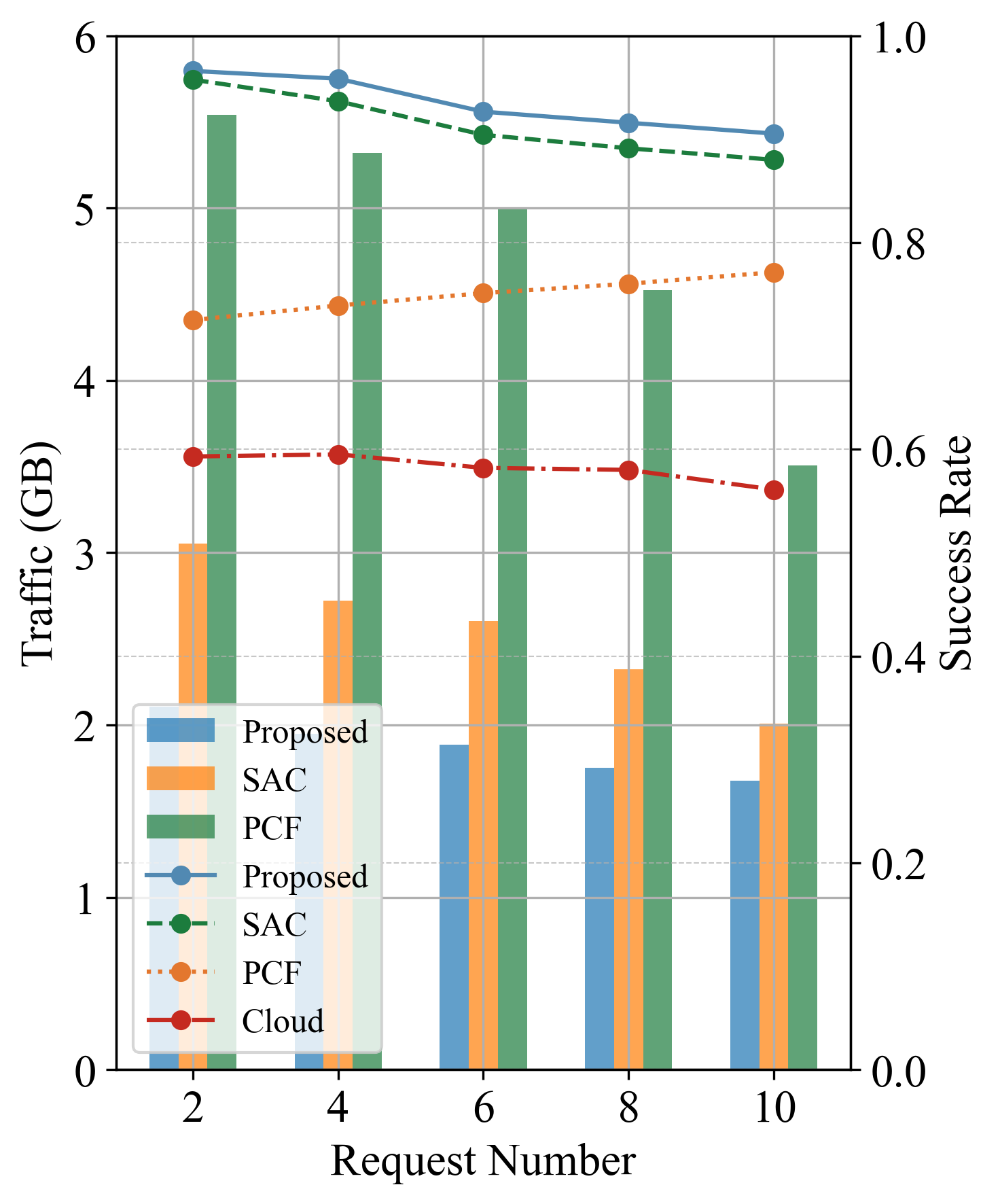}
					\label{Cache_1}
                    \vspace{-5mm} 
			\end{minipage}
		}
	\subfloat[\textnormal{Cache number is 2}] 
	{
			\begin{minipage}[b]{.25\textwidth}
					\centering
					\includegraphics[width=\textwidth]{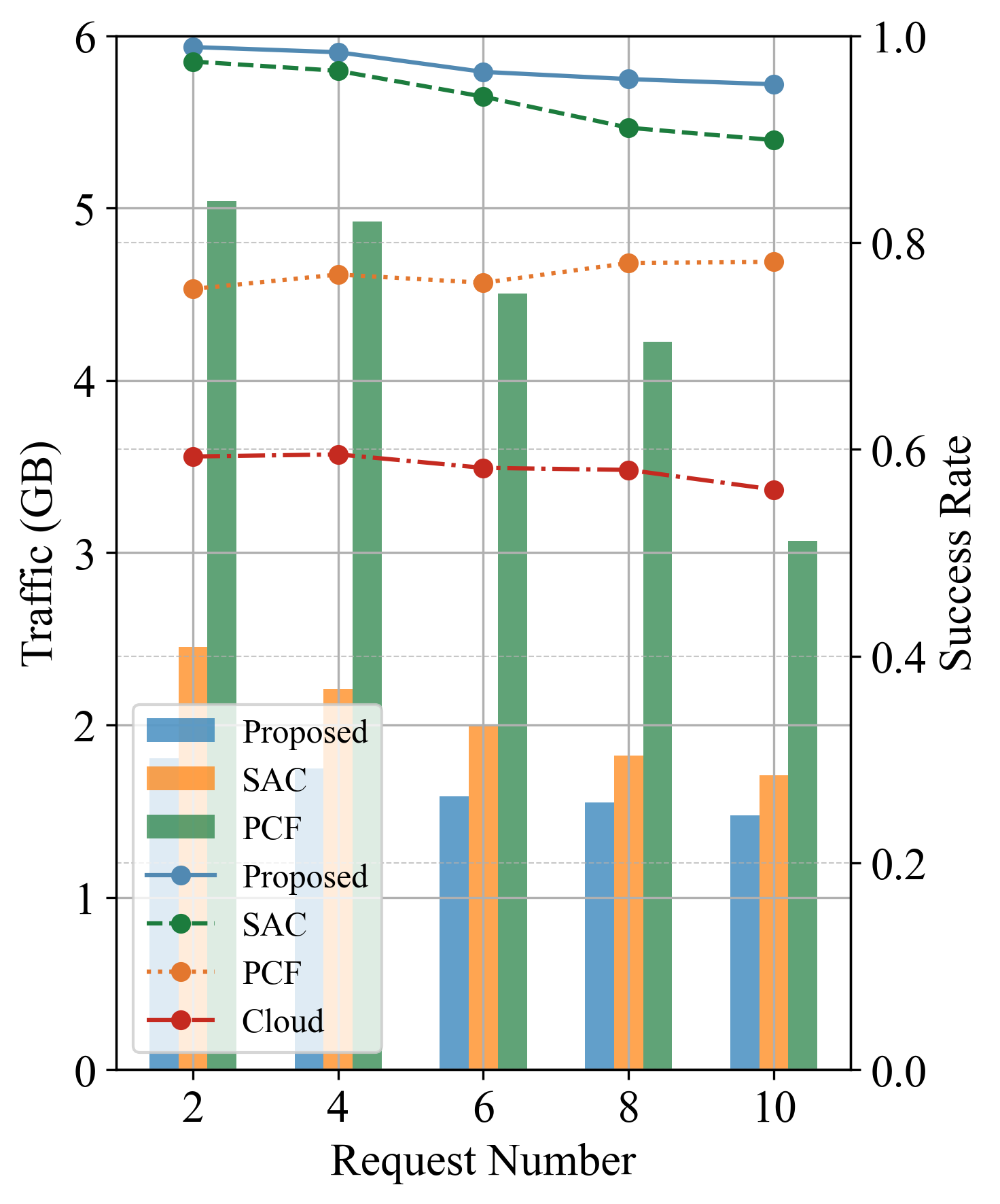}
					\label{Cache_2}
                    \vspace{-5mm} 
			\end{minipage}
		}

	\caption{Performance comparison under different schemes.}
    \label{Different_Cache}
\end{figure}

In Fig.\ref{Different_Cache} we discuss the success rate and traffic performance of four schemes under different cache numbers and different request number per LEO satellite. Fig.\ref{Cache_1} shows the comparison of success rate and update traffic when LEO satellite can only cache one content.  As the number of request increases, the update traffic decreases, which is speculated that the increase in requests reduces the possibility of accidental updates, making the cached content more fixed. The update traffic graph of the proposed scheme is lower than that of the SAC scheme. This is mainly due to the optimized design of the content routing strategy. When the number of requests per LEO satellite is 6, the update traffic of the proposed scheme is approximately 27.6\% and 62.3\% lower than that of SAC and PCF schemes. In terms of success rate, as the number of requests increases, the success rate of the PCF gradually increase. This may be because as the number of requests increases, the probability of requesting content gradually approaches the popularity of the content, which is good for PCF. When the number of requests per LEO satellite is 6, the success rate of the proposed scheme is improved by approximately 59.2\%, 23.3\%, and 2.5\% compared to the cloud, PCF, and SAC schemes, respectively. 

Fig.\ref{Cache_2} shows the the comparison when LEO satellite can cache two contents. Caching two contents results in lower update traffic, because the LEO satellite does not have to update the content as frequently when caching more content. When the number of requests per LEO satellite is 6, the update traffic of the proposed scheme is approximately 20.8\% and 64.8\% lower than that of SAC and PCF schemes. In terms of of success rate, the cloud scheme has the same success rate because it does not retrieve contents from onboard cache. Compared to caching one contents, other three schemes have higher success rate due to more contents can be accessed on satellites. When the number of requests per LEO satellite is 6, the success rate of the proposed scheme is improved by approximately 65.8\%, 26.8\%, and 2.6\% compared to the cloud, PCF, and SAC schemes, respectively. 

\section{Conclusions}\label{SectionV}
In this work, we propose a learning-based framework for joint caching and routing in STECN. By modeling the network as a dynamic spatiotemporal graph, we leverage GNN to extract structural and communication-aware representations for routing. The caching decisions are trained and optimized through SAC. Simulation results demonstrate that our method significantly improves the delivery success rate and reduces traffic cost, highlighting its effectiveness.

\newpage

\end{document}